
\def\service{S}
\catcode`\@=11
\def\unredoffs{\voffset=11mm \hoffset=0.5mm}

%
\newbox\leftpage \newdimen\fullhsize \newdimen\hstitle \newdimen\hsbody
\newdimen\hdim
\tolerance=400\pretolerance=800
%
%
\newif\ifsmall \smallfalse
\newif\ifdraft \draftfalse
\newif\iffrench \frenchfalse
\newif\ifeqnumerosimple \eqnumerosimplefalse
\nopagenumbers
\headline={\ifnum\pageno=1\hfill\else\hfil{\headrm\folio}\hfil\fi}
\def\draftstart{
\magnification=1200 \unredoffs\hsize=130mm\vsize=190mm
\hsbody=\hsize \hstitle=\hsize 
\nolabels
\iffrench
\dicof
\else
\dicoa
\fi
}

\font\elevrm=cmr9

\newdimen\chapskip
\font\twbf=cmssbx10 scaled 1200
\font\ssbx=cmssbx10

\font\caprm=cmr9
\font\capit=cmti9
\font\capbf=cmbx9
\font\capsl=cmsl9
\font\capmi=cmmi9
\font\capex=cmex9
\font\capsy=cmsy9
\chapskip=17.5mm
\def\makeheadline{\vbox to 0pt{\vskip-22.5pt
\line{\vbox to8.5pt{}\the\headline}\vss}\nointerlineskip}
\font\tbfi=cmmib10
\font\tenbi=cmmib7
\font\fivebi=cmmib5
\textfont4=\tbfi
\scriptfont4=\tenbi
\scriptscriptfont4=\fivebi
\font\headrm=cmr10

\font\eightrm=cmr6
\font\sixrm=cmr5
\font\eightmi=cmmi6
\font\sixmi=cmmi5
\font\eightsy=cmsy6
\font\sixsy=cmsy5
\font\eightbf=cmbx6
\font\sixbf=cmbx5
\skewchar\capmi='177 \skewchar\eightmi='177 \skewchar\sixmi='177
\skewchar\capsy='60 \skewchar\eightsy='60 \skewchar\sixsy='60

\def\elevenpoint{
\textfont0=\caprm \scriptfont0=\eightrm \scriptscriptfont0=\sixrm
\def\rm{\fam0\caprm}
\textfont1=\capmi \scriptfont1=\eightmi \scriptscriptfont1=\sixmi
\textfont2=\capsy \scriptfont2=\eightsy \scriptscriptfont2=\sixsy
\textfont3=\capex \scriptfont3=\capex \scriptscriptfont3=\capex
\textfont\itfam=\capit \def\it{\fam\itfam\capit} 
\textfont\slfam=\capsl  \def\sl{\fam\slfam\capsl} 
\textfont\bffam=\capbf \scriptfont\bffam=\eightbf
\scriptscriptfont\bffam=\sixbf
\def\bf{\fam\bffam\capbf} 
\textfont4=\tbfi \scriptfont4=\tenbi \scriptscriptfont4=\tenbi
\normalbaselineskip=13pt
\setbox\strutbox=\hbox{\vrule height9.5pt depth3.9pt width0pt}
\let\big=\elevenbig \normalbaselines \rm}

\catcode`\@=11

\font\tenmsa=msam10
\font\sevenmsa=msam7
\font\fivemsa=msam5
\font\tenmsb=msbm10
\font\sevenmsb=msbm7
\font\fivemsb=msbm5
\newfam\msafam
\newfam\msbfam
\textfont\msafam=\tenmsa  \scriptfont\msafam=\sevenmsa
  \scriptscriptfont\msafam=\fivemsa
\textfont\msbfam=\tenmsb  \scriptfont\msbfam=\sevenmsb
  \scriptscriptfont\msbfam=\fivemsb

\def\hexnumber@#1{\ifcase#1 0\or1\or2\or3\or4\or5\or6\or7\or8\or9\or
	A\or B\or C\or D\or E\or F\fi }

\font\teneuf=eufm10
\font\seveneuf=eufm7
\font\fiveeuf=eufm5
\newfam\euffam
\textfont\euffam=\teneuf
\scriptfont\euffam=\seveneuf
\scriptscriptfont\euffam=\fiveeuf
\def\frak{\ifmmode\let\next\frak@\else
 \def\next{\Err@{Use \string\frak\space only in math mode}}\fi\next}
\def\goth{\ifmmode\let\next\frak@\else
 \def\next{\Err@{Use \string\goth\space only in math mode}}\fi\next}
\def\frak@#1{{\frak@@{#1}}}
\def\frak@@#1{\fam\euffam#1}

\edef\msa@{\hexnumber@\msafam}
\edef\msb@{\hexnumber@\msbfam}

\def\Bbb{\ifmmode\let\next\Bbb@\else
 \def\next{\errmessage{Use \string\Bbb\space only in math mode}}\fi\next}
\def\Bbb@#1{{\Bbb@@{#1}}}
\def\Bbb@@#1{\fam\msbfam#1}

\catcode`\@=12
\def\sla#1{\mkern-1.5mu\raise0.4pt\hbox{$\not$}\mkern1.2mu #1\mkern 0.7mu}
\def\Dbar{\mkern-1.5mu\raise0.4pt\hbox{$\not$}\mkern-.1mu {\rm D}\mkern.1mu}
\def\Abar{\mkern1.mu\raise0.4pt\hbox{$\not$}\mkern-1.3mu A\mkern.1mu}
\def\dicof{
\gdef\Resume{RESUME}
\gdef\Toc{Table des mati\`eres}
\gdef\soumisa{Soumis \`a:}
}
\def\dicoa{
\gdef\Resume{ABSTRACT}
\gdef\Toc{Table of Contents}
\gdef\soumisa{Submitted to}
}

\def\uniset{\rlap{\elevrm 1}\kern.15em 1}
\def\bkR{{\rm I\kern-.17em R}}
\def\bkC{{\rm \kern.24em
            \vrule width.05em height1.4ex depth-.05ex
            \kern-.26em C}}

\def\frac#1#2{{\textstyle{#1\over#2}}}

\def\leaderfill{\leaders\hbox to 1em{\hss.\hss}\hfill}
\def\saclay{\if S\service \spec \else \spht \fi}
\def\spht{
\centerline{Service de Physique Th\'eorique, CEA-Saclay}
\centerline{F-91191 Gif-sur-Yvette Cedex, FRANCE}}
\def\spec{
\centerline{Service de Physique de l'Etat Condens\'e, CEA-Saclay}
\centerline{F-91191 Gif-sur-Yvette Cedex, FRANCE}}
\def\logo{
\if S\service 
\font\sstw=cmss10 scaled 1200
\font\ssx=cmss8
\vtop{\hsize 9cm
{\sstw {\twbf P}hysique de l'{\twbf E}tat {\twbf C}ondens\'e \par}
\ssx SPEC -- DRECAM -- DSM\par
\vskip 0.5mm
\sstw CEA -- Saclay \par
}
\else 
\vtop{\hsize 9cm
}
\fi }
\catcode`\@=11
\def\deqalignno#1{\displ@y\tabskip\centering \halign to
\displaywidth{\hfil$\displaystyle{##}$\tabskip0pt&$\displaystyle{{}##}$
\hfil\tabskip0pt &\quad
\hfil$\displaystyle{##}$\tabskip0pt&$\displaystyle{{}##}$
\hfil\tabskip\centering& \llap{$##$}\tabskip0pt \crcr #1 \crcr}}
\def\deqalign#1{\null\,\vcenter{\openup\jot\m@th\ialign{
\strut\hfil$\displaystyle{##}$&$\displaystyle{{}##}$\hfil
&&\quad\strut\hfil$\displaystyle{##}$&$\displaystyle{{}##}$
\hfil\crcr#1\crcr}}\,}
\openin 1=\jobname.sym
\ifeof 1\closein1\message{<< (\jobname.sym DOES NOT EXIST) >>}\else%
\input\jobname.sym\closein 1\fi
\newcount\nosection
\newcount\nosubsection
\newcount\neqno
\newcount\notenumber
\newcount\figno
\newcount\tabno
\def\content{\jobname.toc}
\def\symbols{\jobname.sym}
\newwrite\toc
\newwrite\sym
\def\authorname#1{\centerline{\bf #1}\smallskip}
\def\address#1{ #1\medskip}
\newdimen\hulp
\def\maketitle#1{
\edef\oneliner##1{\centerline{##1}}
\edef\twoliner##1{\vbox{\parindent=0pt\leftskip=0pt plus 1fill\rightskip=0pt
plus 1fill
                     \parfillskip=0pt\relax##1}}
\setbox0=\vbox{#1}\hulp=0.5\hsize
                 \ifdim\wd0<\hulp\oneliner{#1}\else
                 \twoliner{#1}\fi}
\def\pacs#1{{\bf PACS numbers:} #1\par}
\def\submitted#1{{\it {\soumisa} #1}\par}
\def\title#1{\gdef\titlename{#1}
\maketitle{
\twbf
{\titlename}}
\vskip3truemm\vfill
\nosection=0
\neqno=0
\notenumber=0
\figno=1
\tabno=1
\def\prefix{}
\def\eqprefix{}
\mark{\the\nosection}
\message{#1}
\immediate\openout\sym=\symbols
}
\def\preprint#1{\vglue-10mm
\line{ \logo \hfill {#1} }\vglue 20mm\vfill}
\def\abstract{\vfill\centerline{\Resume} \smallskip \begingroup\narrower
\elevenpoint\baselineskip10pt}
\def\endabstract{\par\endgroup \bigskip}
\def\mktoc{\centerline{\bf \Toc} \medskip\caprm
\parindent=2em
\openin 1=\jobname.toc
\ifeof 1\closein1\message{<< (\jobname.toc DOES NOT EXIST. TeX again)>>}%
\else\input\jobname.toc\closein 1\fi
 \bigskip}
\def\section#1\par{\vskip0pt plus.1\vsize\penalty-100\vskip0pt plus-.1
\vsize\bigskip\vskip\parskip
\message{ #1}
\ifnum\nosection=0\immediate\openout\toc=\content%
\edef\ecrire{\write\toc{\par\noindent{\ssbx\ \titlename}
\string\leaderfill{\noexpand\number\pageno}}}\ecrire\fi
\advance\nosection by 1\nosubsection=0
\ifeqnumerosimple
\else \xdef\eqprefix{\prefix\the\nosection.}\neqno=0\fi
\vbox{\noindent\bf\prefix\the\nosection\ #1}
\mark{\the\nosection}\bigskip\noindent
\xdef\ecrire{\write\toc{\string\par\string\item{\prefix\the\nosection}
#1
\string\leaderfill {\noexpand\number\pageno}}}\ecrire}

\def\appendix#1#2\par{\bigbreak\nosection=0
\notenumber=0
\neqno=0
\def\prefix{A}
\mark{\the\nosection}
\message{\appendixname}
\leftline{\ssbx APPENDIX}
\leftline{\ssbx\uppercase\expandafter{#1}}
\leftline{\ssbx\uppercase\expandafter{#2}}
\bigskip\noindent\nonfrenchspacing
\edef\ecrire{\write\toc{\par\noindent{{\ssbx A}\
{\ssbx#1\ #2}}\string\leaderfill{\noexpand\number\pageno}}}\ecrire}%

\def\subsection#1\par {\vskip0pt plus.05\vsize\penalty-100\vskip0pt
plus-.05\vsize\bigskip\vskip\parskip\advance\nosubsection by 1
\vbox{\noindent\it\prefix\the\nosection.\the\nosubsection\
\it #1}\smallskip\noindent
\edef\ecrire{\write\toc{\string\par\string\itemitem
{\prefix\the\nosection.\the\nosubsection} {#1}
\string\leaderfill{\noexpand\number\pageno}}}\ecrire
}
\def\note #1{\advance\notenumber by 1
\footnote{$^{\the\notenumber}$}{\sevenrm #1}}

\def\nolabels{\def\wrlabel##1{}\def\eqlabel##1{}\def\reflabel##1{}}
\def\writelabels{\def\wrlabel##1{\leavevmode\vadjust{\rlap{\smash%
{\line{{\escapechar=` \hfill\rlap{\sevenrm\hskip.03in\string##1}}}}}}}%
\def\eqlabel##1{{\escapechar-1\rlap{\sevenrm\hskip.05in\string##1}}}%
\def\reflabel##1{\noexpand\llap{\noexpand\sevenrm\string\string\string##1}}}
\global\newcount\refno \global\refno=1
\newwrite\rfile
\def\ref{[\the\refno]\nref}
\def\nref#1{\xdef#1{[\the\refno]}\writedef{#1\leftbracket#1}%
\ifnum\refno=1\immediate\openout\rfile=\jobname.ref\fi
\global\advance\refno by1\chardef\wfile=\rfile\immediate
\write\rfile{\noexpand\item{#1\ }\reflabel{#1\hskip.31in}\pctsign}\findarg}
\def\findarg#1#{\begingroup\obeylines\newlinechar=`\^^M\pass@rg}
{\obeylines\gdef\pass@rg#1{\writ@line\relax #1^^M\hbox{}^^M}%
\gdef\writ@line#1^^M{\expandafter\toks0\expandafter{\striprel@x #1}%
\edef\next{\the\toks0}\ifx\next\em@rk\let\next=\endgroup\else\ifx\next\empty%
\else\immediate\write\wfile{\the\toks0}\fi\let\next=\writ@line\fi\next\relax}}
\def\striprel@x#1{}
\def\em@rk{\hbox{}}

\def\addref#1{\immediate\write\rfile{\noexpand\item{}#1}} 
\def\listrefs{
\ifnum\refno=1 \else
\immediate\closeout\rfile\writestoppt\baselineskip=14pt%
\vskip0pt plus.1\vsize\penalty-100\vskip0pt plus-.1
\vsize\bigskip\vskip\parskip\centerline{{\bf References}}\bigskip%
{\frenchspacing%
\parindent=20pt\escapechar=` \input \jobname.ref\vfill\eject}%
\nonfrenchspacing
\fi}
\def\startrefs#1{\immediate\openout\rfile=\jobname.ref\refno=#1}
\def\xref{\expandafter\xr@f}\def\xr@f[#1]{#1}
\def\refs#1{[\r@fs #1{\hbox{}}]}
\def\r@fs#1{\ifx\und@fined#1\message{reflabel \string#1 is undefined.}%
\xdef#1{(?.?)}\fi \edef\next{#1}\ifx\next\em@rk\def\next{}%
\else\ifx\next#1\xref#1\else#1\fi\let\next=\r@fs\fi\next}
%
\newwrite\lfile
{\escapechar-1\xdef\pctsign{\string\%}\xdef\leftbracket{\string\{}
\xdef\rightbracket{\string\}}}

\def\writestop{\def\writestoppt{\immediate\write\lfile{\string\pageno%
\the\pageno\string\startrefs\leftbracket\the\refno\rightbracket%
\string\def\string\secsym\leftbracket\secsym\rightbracket%
\string\secno\the\secno\string\meqno\the\meqno}\immediate\closeout\lfile}}
\def\writestoppt{}\def\writedef#1{}
\def\eqnn{\global\advance\neqno by 1 \ifinner\relax\else%
\eqno\fi(\eqprefix\the\neqno)}
%
\def\eqnd#1{\global\advance\neqno by 1 \ifinner\relax\else%
\eqno\fi(\eqprefix\the\neqno)\eqlabel#1
{\xdef#1{($\eqprefix\the\neqno$)}}
\edef\ewrite{\write\sym{\string\def\string#1{($\eqprefix%
\the\neqno$)}}%
}\ewrite%
}
%
\def\eqna#1{\wrlabel#1\global\advance\neqno by1
{\xdef #1##1{\hbox{$(\eqprefix\the\neqno##1)$}}}
\edef\ewrite{\write\sym{\string\def\string#1{($\eqprefix%
\the\neqno$)}}%
}\ewrite%
}
\def\em@rk{\hbox{}}
\def\xeqn{\expandafter\xe@n}\def\xe@n(#1){#1}
\def\xeqna#1{\expandafter\xe@na#1}\def\xe@na\hbox#1{\xe@nap #1}
\def\xe@nap$(#1)${\hbox{$#1$}}
\def\eqns#1{(\e@ns #1{\hbox{}})}
\def\e@ns#1{\ifx\und@fined#1\message{eqnlabel \string#1 is undefined.}%
\xdef#1{(?.?)}\fi \edef\next{#1}\ifx\next\em@rk\def\next{}%
\else\ifx\next#1\xeqn#1\else\def\n@xt{#1}\ifx\n@xt\next#1\else\xeqna#1\fi
\fi\let\next=\e@ns\fi\next}
\def\fig{fig.~\the\figno\nfig}
\def\nfig#1{\xdef#1{\the\figno}%
\immediate\write\sym{\string\def\string#1{\the\figno}}%
\global\advance\figno by1}%
\def\xfig{\expandafter\xf@g}\def\xf@g fig.\penalty\@M\ {}%
\def\figs#1{figs.~\f@gs #1{\hbox{}}}%
\def\f@gs#1{\edef\next{#1}\ifx\next\em@rk\def\next{}\else%
\ifx\next#1\xfig #1\else#1\fi\let\next=\f@gs\fi\next}%
\long\def\figure#1#2#3{\midinsert
#2\par
{\elevenpoint
\setbox1=\hbox{#3}
\ifdim\wd1=0pt\centerline{{\bf Figure\ #1}\hskip7.5mm}%
\else\setbox0=\hbox{{\bf Figure #1}\quad#3\hskip7mm}
\ifdim\wd0>\hsize{\narrower\noindent\unhbox0\par}\else\centerline{\box0}\fi
\fi}
\wrlabel#1\par
\endinsert}
\def\tab{table~\uppercase\expandafter{\romannumeral\the\tabno}\ntab}
\def\ntab#1{\xdef#1{\the\tabno}
\immediate\write\sym{\string\def\string#1{\the\tabno}}
\global\advance\tabno by1}
\long\def\table#1#2#3{\topinsert
#2\par
{\elevenpoint
\setbox1=\hbox{#3}
\ifdim\wd1=0pt\centerline{{\bf Table
\uppercase\expandafter{\romannumeral#1}}\hskip7.5mm}%
\else\setbox0=\hbox{{\bf Table
\uppercase\expandafter{\romannumeral#1}}\quad#3\hskip7mm}
\ifdim\wd0>\hsize{\narrower\noindent\unhbox0\par}\else\centerline{\box0}\fi
\fi}
\wrlabel#1\par
\endinsert}
\catcode`@=12
\def\draftend{\immediate\closeout\sym\immediate\closeout\toc
}
\draftstart
\preprint{S93/024}
\title{Conductance statistics in small insulating GaAs:Si wires at low
temperature. II. Experimental study}
\authorname{F. Ladieu}
\address{\saclay}
\authorname{D. Mailly}
\address{\centerline{CNRS-LMM, 196 Ave. H. Ravera, 92220 Bagneux, France}
}
\authorname{M. Sanquer\footnote{$^{1}$}{\sevenrm Email:
sanquer@amoco.saclay.cea.fr}}
\address{\saclay}
\abstract
We have observed reproducible conductance fluctuations at
low temperature in a small GaAs:Si wire driven across the Anderson
transition by the application of a gate voltage. We analyse
quantitatively the log-normal conductance statistics in terms of
truncated quantum fluctuations. Quantum fluctuations due to small
changes of the electron energy (controlled by the gate voltage)
cannot develop fully due to identified geometrical fluctuations of
the resistor network describing the hopping through the sample. \par
The evolution of the fluctuations versus electron energy
and magnetic field shows that the fluctuations are non-ergodic,
except in the critical insulating region of the Anderson
transition, where the localization length is larger than the
distance between Si impurities. \par
The mean magnetoconductance is in good accordance with
simulations based on the Forward-Directed-Paths analysis, i.e. it
saturates to $ {\rm ln} (\sigma (H>1)/\sigma (0))\simeq 1, $ as $ \sigma (0) $
decreases over orders of
magnitude in the strongly localized regime. \par
\endabstract
\vfill
\submitted{J. de Physique I}
\eject
\eject
\input definit.tex
\magnification=1200
\baselineskip=20.0truept
\noindent \overfullrule=0pt \par
\centerline{{\bf CONDUCTANCE STATISTICS IN SMALL INSULATING GaAs:Si WIRES}}
\centerline{{\bf AT LOW TEMPERATURE}}
\centerline{{\bf II: EXPERIMENTAL STUDY}}
\medskip
\centerline{by}
\medskip
\centerline{{\bf F. Ladieu, D. Mailly}$ ^{\dagger} ${\bf\ and M. Sanquer}}
\smallskip
\centerline{{\sl DRECAM/SPEC CE-SACLAY 91191 GIF/YVETTE CEDEX\/}}
\centerline{{\sl and \/}$ ^{\dagger} ${\sl CNRS-LMM 196 Ave H. RAVERA 92220
BAGNEUX\/}}
\centerline{{\sl FRANCE\/}}
\noindent \vglue 3truecm \par
\noindent{\bf ABSTRACT} \par
\smallskip
We have observed reproducible conductance fluctuations at
low temperature in a small GaAs:Si wire driven across the Anderson
transition by the application of a gate voltage. We analyse
quantitatively the log-normal conductance statistics in terms of
truncated quantum fluctuations. Quantum fluctuations due to small
changes of the electron energy (controlled by the gate voltage)
cannot develop fully due to identified geometrical fluctuations of
the resistor network describing the hopping through the sample. \par
The evolution of the fluctuations versus electron energy
and magnetic field shows that the fluctuations are non-ergodic,
except in the critical insulating region of the Anderson
transition, where the localization length is larger than the
distance between Si impurities. \par
The mean magnetoconductance is in good accordance with
simulations based on the Forward-Directed-Paths analysis, i.e. it
saturates to $ {\rm ln} (\sigma (H>1)/\sigma (0))\simeq 1, $ as $ \sigma (0) $
decreases over orders of
magnitude in the strongly localized regime. \par
 \par
\bigskip
\noindent Submitted for publication to: \hfill{SPEC/93-024} \par
\noindent Journal de Physique I \par
\vfill\eject
\noindent{\bf INTRODUCTION} \par
\smallskip
Quantum interferences effects are not well understood in
disordered insulators. This contrasts with the diffusive regime
where their role in the weak localization and Universal
Conductance Fluctuations phenomena has been largely clarified both
theoretically and experimentally $ [1]. $ \par
However, huge, reproducible conductance fluctuations have
been observed for instance in the hopping regime of small
Si:MOSFET $ [2] $ and in lightly doped GaAs:Si samples $ [3]; $ the
conductance statistics are found to be very broad, giving rise to
very high conductance and resistance peaks (as compared to the
averaged value) when the Fermi level or the transverse applied
magnetic field are varied. \par
The mechanism of electronic conduction at finite low
temperature in lightly doped semiconductors has been explained by
Mott $ [4]. $ Let us note $ k_BT_0 $ the level spacing on the scale of the
localization length $ \xi . $ At low temperatures the hopping electrons
optimize the cost due to thermal activation between energy levels
of the initial and final impurities states and the tunnelling
term. This results in a hopping length given, {\sl on\/} {\sl average\/}, by
$ r_0= <r_M> ={\xi \over 2}({T_0 \over T})^{{1 \over d+1}}, $ the Mott hopping
length $ (d $ the
dimensionality). The mean energy
difference between the final and initial impurity levels separated
by $ r_0 $ is: $ E_0= {1 \over 2}k_BT^{{1 \over d+1}}_0 T^{{d \over d+1}}. $
At very low temperatures $ r_{0 } $ diverges
and becomes much larger than $ l, $ the distance between impurities.
$ r_M $ is thought to be the phase coherence length in the insulating
regime. The averaged conductance in large macroscopic sample is
given by $ g\sim {\rm exp} - \left({T_0 \over T} \right)^{{1 \over
d+1}}\simeq {\rm exp} \left(-{r_M \over \xi} \right). $ \par
One has to distinguish two
explanations to describe the conductance fluctuations versus electron
energy in the hopping regime of small samples: fluctuations of geometrical
origin
due to a change of the impurity sites visited by the electrons
travelling through the sample $ [5] $ (incoherent mesoscopic
phenomena $ [6]), $ or quantum fluctuations due to interferences
phenomena for a fixed geometry of hopping paths. \par
Firstly, changes in electronic energy could be sufficient
to induce a change of the impurity sites $ i $ and $ j $ between which the
electrons hop. In other words $ r_M $ fluctuates around $ r_0 $ when one
shifts the electron energy. As we will see, the typical energy
range associated with such a change is the Mott energy $ E_0. $ The
quantum tunnelling resistance depends exponentially on the
distance and on the energy separation of these sites $ [7]: $
$$ R_{ij} \sim  {\rm exp} \left({\vert  E_i \vert  + \vert  E_j\vert  + \vert
E_i- E_j \vert \over 2k_BT}+{\vert  r_i - r_j\vert \over \xi} \right) \eqno
(1) $$
$ (E=0 $ corresponds to the Fermi level). Because few impurity
levels are involved during the hopping through a mesoscopic sample
at very low temperatures, the logarithm of the conductance itself
exhibits large fluctuations. The explanation of large fluctuations
versus the applied magnetic field results, in this geometrical
approach, only from Zeeman shifts of energy of the impurity
states $ [2]. $ \par
Secondly, there exists conductance fluctuations emerging
from quantum interference effects for a {\sl fixed\/}{\bf\ }quantum coherent
hop
(fixed locations and energies for the initial and final impurity
states) of typical size $ r_M \gg  l. $ Because of quantum coherence, one
has to consider all the Feynman paths connecting the initial and
final states, consisting of multi-diffusion paths on intermediate
impurities states. At $ T=0 {\rm K} , $ i.e. when the quantum coherence length
is the length of the sample, only these quantum interferences
persist. They can be regarded as fluctuations of $ \xi $ itself. These
fluctuations are influenced by phase shifts induced by an applied
magnetic flux. \par
Two models have been proposed to take into account the
interference effects in the hopping regime. \par
The first approach, referred to as Forward Directed Path
analysis (FDP), neglects explicitly the quantum interferences
between returning loops due to backward scattering $ [8]. $ This
approach is a perturbative treatment of the deeply localized
electronic states by the intermediate scattering during the
hopping. A crucial assumption is that the localization length is
smaller than the distance between impurities (which is itself
much smaller than the hopping distance). In this situation,
referred to in the rest of this paper as the regime of strong
localization, one has to consider interferences between
Feynman paths of steps $ \sim l, $ $ l $ being smaller than $ r_M. $ As
suggested
first by Nguyen, Spivak and Schklovskii (NSS) $ [8], $ only the
shortest paths -\nobreak\ the Forward Directed Paths\nobreak\ - are important,
because the amplitude of transmission along a $ Nl $ long path is
affected by a prefactor $ {\rm exp} ({-Nl \over \xi} ) \ll  1 $ $ ({l \over
\xi}  > 1), $ exponentially
decreasing with $ N. $ So the Forward Directed Paths approaches are
well adapted at least to the strongly localized regime. The
hypothesis $ \xi  < l $ excludes the critical insulating regime described
by the scaling theory of the Anderson transition. \par
The second approach is based on a Random Matrix Theory
(RMT) applied to the transfer matrix of either conductors or
insulators $ [9]. $ In this global approach resonances as well as
quantum interferences between all sorts of Feynman paths are a
priori included. To some extent, this theory indicates that
returning loops inside the localization domain are essential, and
thus is well adapted to the critical regime of the Anderson
transition, where $ \xi  \gg  l, $ i.e. when electrons are localized over
many impurities sites. \par
FDP and RMT predictions differ drastically for strong
spin-orbit scattering or for the effect of a magnetic field. \par
The FDP approach predicts the existence of a large
positive mean magnetoconductance, which is not the consequence of
interferences between Time Reversal conjugated returning loops
(they are neglected). The mean magnetoconductance $ < {\rm ln} ({g(H) \over
g(0)})\allowbreak > $
depends only on $ r_M $ $ [10], $ and is always positive whatever the
spin-orbit scattering strength. The FDP approaches also predict
large log-normal conductance fluctuations which are smaller versus
the magnetic field than versus the disorder configuration
(non-ergodicity) $ [8]. $ Quantitatively, the amplitude of the
fluctuations $ {\rm var(ln} (g)) $ versus disorder is given by $ [11]: $
$ {\rm var(ln} (g)) \sim  r^{2\omega}_ M $ with $ \omega  = {1 \over 3}\ (
{\rm resp} .{1 \over 5}) $ for $ d=2\ ( {\rm resp} .3). $ \par
By contrast with the FDP approach, the basic symmetries,
as the Time Reversal and Spin Rotation symmetry, are just the
essential ingredients in the Random Matrix Theory. This approach
gives exact results only in quasi-1d geometry, and its
implications have to be weakened in higher dimensions.
Nevertheless numerical simulations in 2d and 3d samples, as well
as previous experiments, yield conclusions which are similar to
some extent to exact RMT results $ [12]. $ Moreover similar conclusions are
obtained in $ d=1,2,3 $ on a completely different model in $ [13]. $ The
main predictions of the RMT approach are that the breaking of the
time reversal symmetry induces changes in the localization length
$ \xi , $ and consequently an exponential magnetoconductance $ [12]. $ The
sign of this magnetoconductance depends critically on the
spin-orbit scattering strength, going from positive to negative
when the spin-orbit scattering increases. This theory also
predicts log-normal fluctuations but with a variance of the
logarithm of the conductance which is related to the mean of the
logarithm of conductance (this is a {\sl one\/} parameter theory):
$ {\rm var(ln} (g)) = -< {\rm ln} (g)> \sim  {L \over \xi} $ $ [9]. $ Note
that contrary to the FDP
result, the fluctuation amplitude -\nobreak\ as well as the mean
magnetoconductance\nobreak\ - depends on $ L/\xi , $ and not only on $ L $ $
(L\sim r_M $ at
finite temperature). The fluctuation is ergodic versus the
magnetic field and the disorder $ [14]. $ \par
It is the aim of this work to test experimentally the
validity domain of both approaches, by addressing the mean
magnetoconductance effect, the distribution of the conductance
fluctuations and the ergodicity $ .A $ submicronic disordered GaAs:Si
wire is driven across the metal-insulator transition by
application of a gate voltage. The conductance of the wire is
measured over many orders of magnitude from the diffusive regime
to the strongly localized regime at very low temperature. To some
extent our observations are similar to previous reported results
$ [2-3], $ but sample, analysis and interpretations differ noticeably. \par
This paper is organized as follows: in the first part we
describe our sample and the vicinity of the metal-insulator
transition when the gate voltage $ V_G $ is varied. This part
includes weak localization fits in the diffusive regime, which
permit the determination of $ L_\varphi = \sqrt{ D\tau_ \varphi} , $ the phase
coherence length
and the effective width of the wire $ (D $ is the diffusion constant,
$ \tau_ \varphi $ the phase-breaking time). The rest of the paper is devoted
to
the insulating regime{\bf .} \par
First, we study the temperature dependence of the
conductance. We show that, because of the one dimensional geometry
of our sample, its behavior with temperature is never given by
the\nobreak\ usual standard Mott's law. Indeed, we explain that
fluctuations of the hopping length around $ r_0 $ cannot be neglected.
The conductance of our sample in the strongly localized regime is
dominated by an exponentially small conductance corresponding to a
hopping distance much larger than the mean Mott's hopping length
$ r_0= < r_M>. $ These considerations are important to explain some
striking experimental observations. \par
We then turn to the study of the lognormal conductance
fluctuations themselves. Those induced by varying the chemical
potential are shown to result from a subtle interplay between
geometrical and quantum fluctuations (\lq\lq Truncated Quantum
Fluctuations\rq\rq , $ [15]). $ Since quantum fluctuations cannot develop
fully as the Fermi energy shifts, we turn to the study of
fluctuations induced by the application of a magnetic field $ H $ and
show that they are of quantum origin. Ergodicity and mean
magnetoconductance behaviors change with the proximity of the
Metal-Insulator Transition, and this permits to clarify validity
domains of FDP and RMT approaches. \par
\medskip
\noindent{\bf I\nobreak\ -\nobreak\ THE METAL-INSULATOR TRANSITION IN OUR
MESOSCOPIC WIRE} \par
\smallskip
{\bf I-1\nobreak\ Sample and Experiment} \par
\smallskip
The sample is a standard Hall bar, with a distance between
successive arms of $ 3 {\rm \mu m} , $ obtained by etching of a Si-doped GaAs
layer. The layer is $ 400 {\rm nm} $ thick grown by Molecular Beam Epitaxy
with a Si concentration of $ 10^{23}m^{-3} $ on a GaAs semi-insulator
substrate. Electron Beam Lithography has been used to pattern the
sample. The subsequent mask was used to etch the active layer
using $ 250V $ argon ions. The width of the sample is approximately
$ 400 {\rm nm.} $ A $ 100 {\rm nm} $ thick aluminium gate has been evaporated
on the
Hall bar. \par
The sample is placed in the plastic mixing chamber of a
compact home-made dilution refrigerator. For electrical
measurements, coaxial cables are used between $ 300 {\rm K} $ and $ 4 {\rm K}
, $ and
strip lines between $ 4 {\rm K} $ and the mixing chamber. All the lines are
properly filtered. The resistance is obtained by measuring the
current passing through the sample with a Keithley 617
electrometer. The controlled excitation voltage supplied by the
electrometer is divided, and the I-V nonlinearities have been
precisely studied (see later). The electrometer is controlled by
computer, and each measurement cycle consists of 10 voltage
inversions followed by a 3\nobreak\ sec waiting time and 6 measurements
(conversion time 0.3\nobreak\ sec.). So the resistance results from an
average of 60 measurements. The offset voltage is approximately
$ 100\mu V $ for very different measured resistances. We have not
detected any offset current. \par
At very low temperature in mesoscopic samples, one has to
be very careful about excitation and offset voltages applied
across the sample $ [1]. $ A common problem is to measure large
resistances with small enough excitation voltages to be in the
linear I-V regime. Figure 1 shows a typical I-V curve obtained at
$ T = 91 {\rm mK} $ in our sample. The characteristic is well fitted by:
$$ I = A {\rm sh} ( {V_{ds} + V_{ {\rm offset}} \over B} ) {\rm \ \ \ with\ \
\ } A= 4.10^{-12}A, {\rm \ \ \ } B= 5.10^{-4}V \eqno (2) $$
and $ V_{ {\rm offset}}= -2\ 10^{-4}V. $ The conductance is given by:
$$ g = {\partial I \over \partial V}_{ V_{ds}+ V_{ {\rm offset}}= 0}= {A
\over B} = 8.10^{-9}S. \eqno (3) $$
\par
The sh function is the simplest way to introduce the
voltage non linearities; we do not see any rectifying behavior in
our experiment. All the presented results are obtained in the I-V
linear regime. \par
The low-temperature conductance of the sample depends on
the history of the cooling down from room temperature. In other
words the conductance for $ V_g=0V $ depends for instance on whether
the sample has been cooled under $ V_g=+1V  $or under $ V_g=-1V. $ The
conductance is systematically larger in the later case. There
persist long time relaxations at $ T=4 {\rm K} $ after a large variation of
$ V_g. $ A systematic study permits us to conclude that this
relaxation is not due to a dynamic of disorder seen by the
electrons, but to a slow variation of the Fermi level. In fact,
after a large cycling in $ V_G, $ the observed conductance fluctuation
patterns are translated in $ V_g $ but not at all decorrelated. This is
consistent with a retarded response of the number of electrons to
large changes of $ V_g, $ with the disorder configuration unchanged.
One can qualitatively take into account the observed facts by
supposing that the charge configuration of electronic traps inside
the depletion barrier under the gate is not the equilibium
configuration corresponding to the nominal $ V_g $ at low temperature.
The difference results from the slow kinetics of trapping and
release processes for the electrons at low temperature. The charge
configuration in the depletion layer influences the number of
electrons and the Fermi energy in the center of the wire. \par
These relaxations can be avoided by restricting the range
of gate voltage changes in a given experiment at low temperature,
or if not possible, by varying the gate voltage back and forth a
few times in the corresponding range before the experiment. With
the help of these experimental procedures the conductance pattern
is fully reproducible as long as the sample is kept below $ T=4 {\rm K} . $
\par
\smallskip
{\bf I-2\nobreak\ The diffusive regime} \par
\smallskip
Figure 2 shows the magnetoconductance observed at low
temperature for a large gate voltage $ V_G, $ such as the conductance
of the wire is relatively large. For this value of $ V_g, $ the
temperature dependence of the conductance is weak below $ T=4.2 {\rm K} . $
It is impossible to fit this dependence with a variable range
hopping activation law (as we will do in the insulating regime),
because it gives too small $ T_0 $ parameters (for instance
$ T_0\simeq  50 {\rm mK} < T $ for $ V_G=1.8V). $ We fit the mean behavior of
the
large positive magnetoconductance with standard 1D weak
localization formula $ [16] $ and we find $ L_\varphi = 130 {\rm nm} $ and an
effective
cross section $ W^2= (65 {\rm nm} )^2 $ (the sample has been rotated in the
magnetic field and the magnetoconductance is found the same, which
indicates that the cross section is isotrope). The effective
length of the sample is evaluated to be $ 5 {\rm \mu m} , $ because in our
two-probe measurement a part of two thin arms under the gate
contributes to the conductance. The magnetic field $ H_c $ which gives
a flux quantum through $ L_\varphi W $ is $ H_c={h \over e} {1 \over
L_\varphi W} =.42\ {\rm Teslas.} $ This gives
the good order of magnitude for the correlation field of the
magnetoconductance fluctuations. The amplitude of the
fluctuations, if they are supposed to be the Universal Conductance
Fluctuation, is given by $ [17]: $
$ \delta g(H)\simeq{ e^2 \over h} \sqrt{{ 4 \over 15}({L_\varphi \over
L})^3}\simeq 2.2\ 10^{-3}({e^2 \over h}) $ in good accordance with
the experiment. \par
Figure 3 shows the variation of the conductance (in units
of $ {e^2 \over h}) $ as function of the applied gate voltage for $ T=100
{\rm mK} . $ The
conductance exhibits reproducible Gaussian fluctuations as a
function of $ V_G, $ of amplitude similar to the conductance
fluctuations induced by the transverse applied magnetic field, and
so in accordance with the estimation of the Universal Conductance
Fluctuation. \par
In accordance with the scaling theory of the Anderson
transition, we expect that the transition occurs for a conductance
at the phase coherence length of order $ {e^2 \over h}. $ For our sample
consisting approximately of $ {L \over L_\varphi}  \simeq  40 $ quantum boxes
in series, this
criterion corresponds to a conductance of order $ 2.5\ 10^{-2} {e^2 \over h},
$
close to the observed value which separates non-activated and
activated behaviors for the temperature dependence below $ T=4 {\rm K} . $
\par
In this range of conductances, the conductance fluctuation
departs from its value in the diffusive regime, growing and
becoming asymmetric with tails to low conductances. \par
With the estimated effective cross section, and supposing
that the concentration of electrons is close to the critical
concentration in GaAs for the Metal-Insulator Transition
$ n_c=1.6\ 10^{22}m^{-3}, $ we find a mobility of $ \mu  \simeq  3600 {\rm
cm}^2/Vs. $ Close to
the transition, we obtain that $ \lambda_ F\simeq  65 {\rm nm} $ comparable to
the width
of the sample, $ E_F\simeq  45 {\rm K} , $ the elastic mean free path $
l\simeq 24 {\rm nm} $
comparable to the distance between Si atoms, and $ k_Fl \simeq  2. $ \par
\smallskip
{\bf I-3\nobreak\ The Anderson Transition} \par
\smallskip
As the gate voltage is reduced, the number of electrons in
the wire decreases as their Fermi energy:
$$ eN = \int^{ }_{ } C_{ {\rm gate}}(V_g) dV_g \eqno (4) $$
Typically, we estimate that $ C_{ {\rm gate}}\simeq  1.5\ 10^{-16}F $ and we
neglect its
gate voltage dependence. Near the critical Mott's concentration
$ n_c\simeq  1.6\ 10^{22}m^{-3} $ and taking a 3D density of states, we
estimate
that a variation $ \Delta V_g\simeq  10mV $ corresponds to $ \Delta E_F\simeq
1 {\rm K} $ (Note that,
with this crude estimation, the gate voltage range needed to
deplete the wire completely from the $ n_C $ value is $ \simeq 0.5V). $ \par
The Anderson transition takes place below a certain
critical gate voltage, and the temperature dependence of the
conductance becomes activated. This is apparent on figure 4A,
where $ {\rm ln} (G) $ is plotted versus $ T^{{-1 \over 2}} $ for various gate
voltages. An
interesting point is that the activated behavior saturates below a
temperature which increases when the sample becomes more
insulating. We will discuss that saturation in section I-5. \par
In the restricted range of temperature where the Mott
hopping regime is seen: $ g \sim  {\rm exp} \left(- \left({T_0 \over T}
\right)^{{1 \over d+1}} \right) $ $ (d $ the dimensionality),
it is difficult to evaluate precisely the actual value of the
exponent. One first point is that the exponent must give a
reasonable estimation for the parameter $ T_0, $ i.e. it cannot exceed
60\nobreak\ kelvins, the energy of a single Si impurity state in GaAs. For
this reason, one cannot choose an exponent of $ {1 \over 4} $ $ (d=3) $ since
this
would give a $ T_0 $ of order of a thousand $ {\rm K} . $ Moreover, since the
effective cross section of our sample at the M.I.T. is only $ 65 {\rm nm}^2 $
and since it decreases when Vg is diminished, it is not surprising
that, below M.I.T., our sample should be a 1D wire
$ (r_0>W,\ d=1) $. \par
\smallskip
{\bf I-4\nobreak\ The One-Dimensional Hopping Regime} \par
\smallskip
It has been first pointed out by Kurkijarvi $ [18], $ that one
has a simple $  T^{-1} $ activation law for the conductance for a given
1D wire in Mott's regime. This results from the fact that a single
hop dominates the measured resistance. A priori, the slope of this
single activation law only gives the energy activation of the
dominant link and not directly $ T_0, $ the {\sl mean\/} energy spacing on the
scale of the localization domain. We will see later that when
{\sl averaging over disorder\/} is made, one recovers an exponent $ {1 \over
2} $ whose
slope is a function of both the length of wire and of $ T_0. $ Let us
explain why. \par
Qualitatively, let us note that in samples at $ d=2 $ or 3,
Mott's law is observed without averaging over many samples. This
is because when $ d>1 $ self-averaging occurs within each sample,
allowing to consider only a typical resistor $ (r_0,E_0) $ given by
Mott's law to calculate the resistance of the whole sample. But in
1D wires, such an averaging does not take place: since elementary
resistors are added always in series, one has to consider the
strongest one (and not the mean one) in order to evaluate the
resistance of the wire. \par
Such an idea can be quantitatively developed. We now
summarize what comes out of a detailed analysis of the Mott's VRH
in 1D wires $ [5,6,15]. $ Let us consider a long wire without
fluctuations of quantum origin which allows us to use equation
(1)
for each elementary resistance $ R_{ij}  $ and to get their values as
soon as the distribution $ (x_i,E_i) $ of localized states is known. One
statistically neglects resonant or direct tunnelling since we
assume $ L\gg r_0. $ Using an assumption of local optimisation (at
each step the electron chooses the less resistive hop), one can
self consistently solve the problem of 1D hopping $ [15]. $ Due to
possible local lack of levels near the chemical potentiel $ \mu , $
lengths of elementary hops fluctuate around $ r_0, $ giving for $ R_{ij}  $ a
distribution whose width $ w_{ij} $ is so large that the addition of
$ N={L \over r_0} $ resistances $ R_{ij} $ in series does not self-average (as
long
as $ N $ is not extremely large). Note that such a method is
consistent only if $ w_{ij}\gg  w_q, $ where $ w_q $ is the width of the
distribution of resistances due to quantum interferences $ (w_q $ can
be regarded as the fluctuation of $ {1 \over \xi} $ in (1)). \par
One can show that, if $ N<N^\ast ={1 \over a}e^{2(2{r_0 \over \xi} )^2} $ $
(a\simeq 2), $ the
resistance of a wire is entirely dominated by only one elementary
most resistive hop: $ R_{ {\rm max}}= {\rm Max}_N(R_{ij}) $ whose average
value is size
dependent. Estimation of $ R_{ {\rm max}} $ gives:
$$ {\rm ln} \ R\simeq {\rm ln} \ R_{ {\rm max}}={r_{ {\rm max}} \over \xi}
\eqno (5) $$
$$ {\rm with} \ \ \ \ <r_{ {\rm max}}>=2r_0 \sqrt{ 2 {\rm ln} (aN)}=\xi ({T_0
\over T})^{1/2} \sqrt{ 2 {\rm ln} (aN)} \eqno (5 {\rm bis} ) $$
\par
Note that {\sl in average over disorder\/}, one still has a $ T^{-1/2} $
law. The measured lnR does not directly give $ T_0 $ but features of
the dominant hop. Nevertheless $ T_0 $ -\nobreak\ the important averaged
microscopic energy\nobreak\ - can be estimated for our experimental
parameter $ {\rm ln} \ R $ and for reasonnable $ \xi : $ in the companion
paper $ [15] $ a
simulation of our wire for $ {\rm ln} \ R\simeq +9 $ at $ T=.45 {\rm K} $ is
presented with:
$ \xi =2l\simeq 50 {\rm nm} $ and $ T_0=6 {\rm K} $ (see the comments in $
[15] $ on the slight
discrepancy between calculated and measured $ {\rm ln} \ R). $ $ \xi \simeq l,
$ so we
call this regime \lq\lq strongly localized\rq\rq , by contrast with the
\lq\lq barely insulating regime\rq\rq\ that one encounters near M.I.T. where
$ T_0 $ is not large enough compared to $ T $ to allow a description in
terms of variable range hopping. In this regime $ \xi $ must be given in
order of magnitude by $ L_\varphi \simeq 130 {\rm nm} $ (at very low
temperature), i.e.
$ \xi \gg l. $ \par
Moreover, we found numerically that the whole experimental
range of conductances corresponds to variations of $ T_0 $ between $ 2 {\rm K}
$
and $ 10 {\rm K} . $ Let us emphasize that these values are significantly
lower than those naively extracted from data in $ T^{{-1 \over 2}} $scale (see
figure
4A) which, as we explained, is definitely not relevant for a {\sl given\/}
wire in Mott's regime. \par
\smallskip
{\bf I-5\nobreak\ Saturation of the Conductance at low Temperature} \par
\smallskip
As noted before, the temperature dependence of the
conductance exhibits a saturation below a temperature which
increases when the gate voltage decreases. Because all the
measured conductance properties become temperature independent, it
is likely to incriminate electron heating by radiofrequency
voltage sources (let us recall that the conductance is recorded
in the I-V linear regime). Voltage radiofrequency noise is a
priori more efficient to heat electrons when the conductance is
high. However the conductance saturation is clear only when the
conductance is low (small $ V_G). $ Moreover the saturation
temperature is the same for the peaks and the valleys of the
conductance pattern even for peak-to-valley ratio as large as $ 10^2, $
in the strongly localized regime. This is hardly compatible with
simple heating. \par
Even if it is difficult to rule out heating by
radiofrequency pickup, the observed saturation up to $ T=400 {\rm mK} $ seen
in the strongly localized regime could be due to intrinsic
physical effects: either resonant tunnelling processes or the
existence of plateaus in the temperature dependence of a
mesoscopic 1D wire in the hopping regime $ [15]. $ \par
A crossover from hopping at high temperature to $ T
$-independent tunnelling at low temperature should happen if the
diverging Mott hopping length $ r_0 $ (more precisely $ r_{ {\rm max}}) $
becomes of the
order of the sample length at low $ T $ $ [2,19]. $ But the estimation of
$ r_{ {\rm max}}\simeq  600 {\rm nm} $ obtained from the above reported
estimation of $ T_0 $ is about
10 times smaller than our sample length when the saturation of $ g $
occurs. The resonant tunnelling through the sample is negligible under
this condition. Another observation against the resonant tunneling
picture is that the measured conductance is always decreasing when the
temperature decreases, even for sharp conductance peaks. But it is well
known that inelastic processes always decrease the resonant conductance
in the tunnelling processes, whereas phonons always increase the
hopping conductance. For these reasons we do not believe that resonant
or direct tunnelling processes are of importance in our geometry. \par
Apart from resonant tunnelling or heating, special features of
temperature dependence in 1D V.R.H. could give rise to temperature
saturation. As reported in figure 2 of $ [15], $ one has to distinguish two
main cases for the temperature dependance. First, when temperature is
such that values of $ N=L/r_0 $ are large enough (precisely when $ N>N^\ast $
defined above), $ {\rm ln} \ R $ should vary as $ {1 \over T}. $ Since $ r_0 $
diminishes as $ T $
increases, such a regime only arises at quite high temperatures, let us
say: $ T>T^\ast . $ Using the definition of $ N^\ast , $ one can see that $
T^\ast $ increases
when the sample becomes more insulating (i.e. when $ T_0 $ increases). One
can indeed see on figure 4A -\nobreak\ and this is a general trend\nobreak\ -
that
activated behaviour is valid above a temperature which grows as the
sample is driven to a more insulating regime. We estimate, using the
definition of $ N^\ast $ with the experimental parameters, that: $ T^\ast
\simeq  1-2 {\rm K} $ in
the strongly insulating regime. \par
What happens if $ N<N^\ast ? $ As discussed in $ [15] $ we think that
in this case the activation energy of the dominant link can be
very weak, leading to an apparent saturation of $ R $ with decreasing
$ T. $ If this happens, such a non-activated link will remain dominant
as long as the second-dominant activated link becomes more
resistive because of decreasing $ T. $ Thus $ T $-dependence of $ R $ will be
a succession of \lq\lq activated segment\nobreak\ -\nobreak\ apparent
plateau\rq\rq\ and
ref. $ [15] $ shows that in a logarithmic scale of $ T $ plateaus and
segments are of same size. When averaging over many samples, one
should however recover Mott's 1D law due to random location of
segments and plateaus for different samples. \par
However, the observed saturation of $ R $ is larger than the
size of plateaus predicted in $ [15] $ and moreover we never see an
activated segment at temperatures lower than the temperature at
which saturation begins. Therefore, we think that heating by rf
pick-up could be partly responsible for the observed saturation. \par
Up to here, the study of the temperature dependence in the
localised regime has been carried out without taking into account
any quantum fluctuations. We focus now on conductance fluctuations
versus Fermi energy and on the effect of magnetic field, which
will give us much more insight into the relevance of zero
temperature theories for our experiment. \par
\medskip
\noindent{\bf II\nobreak\ -\nobreak\ CONDUCTANCE FLUCTUATIONS IN THE LOCALIZED
REGIME} \par
\smallskip
Figure 5 shows the conductance as function of the gate
voltage (over a large range of $ V_G) $ for two temperatures: $ T=4.2 {\rm K}
$
and $ T=100 {\rm mK} $ (a thermal cycling up to room temperature has been
applied between the two records). The relative fluctuation
becomes enormous for small values of the conductance (exceeding
sometimes two orders of magnitude), so that a semilog
representation is more adapted (figure 6). \par
\smallskip
{\bf II-1\nobreak\ Quantitative analysis of the log-normal conductance
fluctuations} \par
\smallskip
We develop in this section a quantitative analysis of the
log-normal conductance fluctuations, based on the considerations
developed successively by P.A. Lee $ [5], $ Raikh and Ruzin $ [6], $ and
Ladieu and Bouchaud $ [15]. $ \par
Figure 8 shows $ \delta {\rm ln} (g)\ {\rm versus\ } < {\rm ln} (g)> $ for $
T=100 {\rm mK} $ and
$ H=0 $\nobreak\ Teslas. $ < {\rm ln} (g)> $ is obtained by numerical
smoothing of $ {\rm ln} (g) $ to
remove the short $ V_G $-range fluctuations. Two experiments differing
only by a thermal cycling to room temperature are presented in
order to improve the statistics. \par
As we reported in the preceding section, the measured $ {\rm ln} \ R $
is dominated by the most resistive link $ R_{ {\rm max}} $ whose value is size
dependent. Thus amplitude of fluctuations is given by the width
$ w_N $ of $ R_{ {\rm max}} $ distribution. Estimation of $ w_N $ leads to $
w_N\ll w_{ij}, $ and
gives:
$$ {\Delta {\rm ln} \ R \over {\rm ln} \ R}={1 \over 2 {\rm ln} (aN)} \eqno
(6) $$
\par
Fortunately, this prediction depends weakly on the single
adjustable parameter $ N, $ for realistic large values of $ N. $ We
numerically
found (see figure 4 of $ [15]) $ that $ N\simeq 53, $ but even taking $
N=25-100 $
$ (r_0=50-200 {\rm nm} ), $ we get a small dispersion:
$$ {\Delta {\rm ln} \ R \over {\rm ln} \ R}=0.11\pm .015 \eqno (7) $$
This prediction is reported on figure 8, in very good accordance
with the experimental data. \par
Therefore, at this point, one does not need to invoke the
quantum coherence to explain the observed amplitude of $ \delta {\rm ln} (R).
$ We
detail now the arguments which justify the introduction of quantum
fluctuations within the most resistive hop. \par
The predicted energy width for the geometrical
fluctuations is given by $ [6-15]: $
$$ \Delta E_{ {\rm geo}}= 2E_0 {1 \over \sqrt{ 2 {\rm ln} (2N)}} \simeq  1
{\rm K} \eqno (8) $$
typically in our strongly localized regime  (for $ T\simeq .5 {\rm K} ). $
This
energy scale is in fact twice the mean energy spacing of levels
lying within $ r_{ {\rm max}}. $ But, very recently, numerical simulations of
quantum fluctuations versus Fermi energy at $ T=0 {\rm K} $ have been
carried out $ ([20] $ companion paper). They have suggested that the
typical width in energy $ \Delta E_{ {\rm qu}} $ of these fluctuations is of
the order
of the mean energy level spacing within the finite quantum
coherent system. Let us assume, as usually, that quantum coherence
is preserved on the scale of each hop at finite temperature. Then,
we get that quantum interferences in the dominant link change
completely within a scale in energy given by the mean level
spacing within $ r_{ {\rm max}} $ at finite temperature. Therefore, we get
$ \Delta E_{ {\rm qu}}\simeq \Delta E_{ {\rm geo}} $ at finite temperature.
Crudely speaking, this means
that within $ r_{ {\rm max}} $ quantum interferences are dominated by
diffusion
on levels whose energy is the closest to initial and final
energies of hop. This energy is simply $ \simeq \Delta E_{ {\rm geo}}. $ \par
Of course, this latter statement is concerned with only
mean energy scales. Therefore, we think that observed fluctuations
are partly of quantum origin, depending on each particular
fluctuation: if for a given hop, quantum interferences change with
energy more quickly than geometrical fluctuations, then the
fluctuation will be of quantum origin and therefore $ T $ independent.
If the inverse situation takes place we will get a strongly
$ T $
dependent fluctuation just given by geometrical considerations.
Indeed, even for a given hop, providing that the value of the
resistance is given exclusively by equation (1), the fluctuation induced
by varying Fermi energy is very sensitive to any shift of
temperature. \par
Figure 4B gives an example of a fluctuation of quantum
origin. Indeed, one can see that this conductance fluctuation $ \delta {\rm
ln\ } g $
exhibits no or very weak temperature dependence, even in a
temperature range where the mean conductance keeps on decreasing
with decreasing $ T $ (here, e.g. between $ T=1 {\rm K} $ and $ T=400 {\rm mK}
). $ This
behaviour suggests that finite temperature models totally removing
quantum interferences are incomplete. \par
Now, let us consider the amplitude of a quantum fluctuation on
the dominant resistor. It is worth noting that the zero temperature RMT
or FDP approaches predict: $ \Delta {\rm ln\ } R\simeq ({T_0 \over T})^\alpha
>1 $ (see $ [15], $ $ \alpha =1/4 $ or 1/10
respectively for R.M.T. and F.D.P.), whereas the geometrical one is
always $ \losim 1 $ in our experiment. This quantitative analysis shows that
the
fluctuation that we observe cannot be the full quantum one, but is
truncated by the geometrical fluctuation. This means that when a
quantum fluctuation inside the largest (dominating) resistor yields a
large increase of the resistance, the electrons hop to a different
final impurity site. On the contrary, when a quantum fluctuation yields
a large decrease of the resistance, the second largest resistor starts
playing a leading role, therefore limiting again the fluctuation of
measured $ {\rm Ln\ } R. $ \par
Near the Anderson transition, $ w_{ij} $ is no longer much larger than
the estimated quantum fluctuations $ [15], $ which means that the above
considerations break-down since the method used is no longer valid.
Physically, this means that the effect of interferences within $ \xi $ itself
can no longer be ignored (quantum fluctuations can be regarded as
fluctuations of $ \xi ). $ Moreover because $ w_{ij} $ decreases, the whole
conductance is less and less controlled by the weakest link. In this
regime, the quantum fluctuation should develop fully, but this range is
too narrow to allow a quantitative test. Moreover, the temperature
dependence of fluctuations in this regime is much more marked than in
the regime of figure 4B. This emphasizes that the description of the
vicinity of the transition requires a model where quantum
fluctuations are fully taken into account, and not only considered
on the dominant link. \par
The study of fluctuations versus the Fermi energy shows the
subtle interplay between quantum and geometrical fluctuations. The
application of a magnetic field can induce Zeeman shifts of energy
levels $ E_i $ in (1), and consequently induce geometrical fluctuations. On
the other hand magnetic flux can change the quantum interferences and
induce quantum fluctuations. We will see in the next section that
magnetoconductance fluctuations are purely due to quantum interference
effect in our sample. \par
\smallskip
{\bf II-2\nobreak\ The Fluctuations in Applied Magnetic Field versus the
Fluctuations in }$ {\bf V}_ {\bf G} $ \par
\smallskip
Figure 8 presents a detail of the conductance fluctuation
versus gate voltage and applied magnetic field for both very low
and moderately low conductances at $ T=100 {\rm mK} $ (see figure 6). \par
\smallskip
{\sl II-2A\nobreak\ Strongly localized regime: Non ergodicity\/} \par
For the very low conductances in a linear scale
representation, conductance peaks seem to appear just by
application of the magnetic field, as in reference $ [2]. $ In a
logarithmic representation, however, such conductance peaks
correspond to maxima of the conductance in zero field. Moreover,
the applied magnetic field is unable to decorrelate the pattern of
the conductance fluctuations versus the gate voltage. This
situation is precisely referred to as non-ergodic $ [3]: $
$$ {\rm var} ( {\rm ln\ } R)_H\simeq  0.22< {\rm var} ( {\rm ln\ }
R)_{V_G}\simeq  1.10 \eqno (9) $$
\par
This is not, strictly speaking, a proof that there is
non-ergodicity in this strongly localized situation, because one
first has to know if the field scale appearing in the problem is
not too large, or, equivalently, if the statistics over the
magnetic field is complete. Our experimental field range is
limited below 4 or 5\nobreak\ Teslas because of the large negative mean
magnetoconductance associated with the shrinking of atomic
orbitals for higher field $ [21]. $ In fact when the condition:
$ H\gg{ \hbar \over e}{1 \over al}\simeq 3\ {\rm Teslas} $ (for $ a=a_{ {\rm
Bohr}} $ and $ l=20 {\rm nm,} $ the distance
between Silicon impurities) is fullfilled, the magnetic field
modifies the shape of each wave function, and not only the phase
along the Feynman paths. So we restrict ourselves to the low field
range for interpretation. \par
Between 0 and 3.4\nobreak\ Teslas, typically we only see 2 or 3
oscillations of $ {\rm ln} (g(H)) $ in the strongly localized regime. The
correlation field (difficult to be estimated) is of order
1\nobreak\ Tesla (a quantum of flux $ {h \over e} $ is put through $ (64 {\rm
nm} )^2 $ for 1\nobreak\ Tesla).
Nevertheless, the comparison with the barely localized situation
shows that the experiment distinguishes, in practice, the ergodic
and non-ergodic cases\nobreak\ -\nobreak\ even for one or two oscillations of
magnetoconductance. \par
The observed non-ergodicity implies that the magnetic
field is unable to induce geometrical fluctuations. On the
contrary, a strong Zeeman shift would change all the impurity
energies and thus the geometry of hopping paths $ [2-3] $ inducing
geometrical fluctuations. We do not see the magnetic field
translating the maxima of $ {\rm ln} \ g $ $ [2], $ and so Zeeman effects are
negligible in our
sample for our field range. \par
The experiment shows that the quantum fluctuation versus
magnetic field $ (\Delta {\rm ln\ } R_H<1), $ is smaller than the geometrical
fluctuation $ (\Delta {\rm ln\ } R_{ {\rm geo}}\simeq  1). $ This is in the
spirit of the Nguyen,
Spivak and Shklovskii model $ [8], $ where the quantum fluctuation
versus magnetic flux is smaller than any other kind of
fluctuation. This has been already noticed by Orlov et al. in
reference $ [3]. $ Furthermore we have suggested in section II-1 that the
geometrical fluctuation is smaller than the quantum fluctuation
versus energy $ (\Delta {\rm ln\ } R_{ {\rm qu}}> 1) $ (\lq\lq truncated
quantum fluctuation\rq\rq ).
This allows us to conclude that the quantum fluctuation is larger
versus energy than versus magnetic field. To our knowledge, there
is no attempt to modelize the fluctuation versus energy at $ T=0 {\rm K} $
except the work of Avishai and Pichard $ [20]. $ In the strongly
localized regime, their numerical results show a similar
non-ergodic behavior, precisely when standart RMT results start to
fail. \par
\smallskip
{\sl II-2B\nobreak\ Barely localized regime: Ergodicity\/} \par
Figure 8D shows the conductance as function of $ V_G $ and applied
magnetic field at $ T=70 {\rm mK} $ for a range of conductance just on the
insulating side of the Anderson transition: typically
$ < {\rm ln} (g(H=0))>\sim -5(g\sim 7\ 10^{-3}), $ whereas the transition
takes place
for $ < {\rm ln} (g(H=0))>\sim -3.7\ (g\sim 2.5\ 10^{-2}). $ For these
conductances,
$ T_0\simeq  2 {\rm K,} $ so that we are in the limiting case of the VRH
regime. In this
range of conductance, the shape of the fluctuations is reminiscent of
what is observed more deeply in the insulating regime. From the
experiment it will be pointless to argue any further about the exact
position of the transition. \par
By contrast to the strongly insulating regime, near the
Anderson transition, the experiment indicates the validity of the
ergodic hypothesis\nobreak\ -\nobreak\ formulated first in the diffusive
regime for small
disorder parameter $ (k_Fl)^{-1}: $
$$ {\rm var} ( {\rm ln\ } R)_H\simeq 0.19\simeq {\rm var} ( {\rm ln\ }
R)_{V_G}\simeq 0.27 \eqno (10) $$
Our experiment shows that it is still valid at least very close to
the transition, and by continuity in the critical insulating regime. We
have seen that near the Anderson transition it is no longer relevant to
separate geometrical and quantum fluctuations: the analysis performed
in the strongly localized regime fails as already mentioned in section
II-1. \par
Because the estimated $ \xi $ becomes quite large with respect to the
distance between impurities, the electrons are no longer fixed to a
given impurity but localized in shallow regions, which are changed by
application of a magnetic field. Because of this redistribution, the
ergodic hypothesis is realistic. It is indeed numerically obtained by
Avishai and Pichard near the Anderson transition $ [20]. $ \par
But as we mentioned in II-1, the extension of this critical
regime $ (\xi \gg l) $ in our MBE grown GaAs:Si sample appears to be quite
narrow. We believe that it is much more developed in less pure samples
like amorphous alloys. Because of this narrowness, it is difficult to
be more quantitative. \par
In both NSS model and RMT model, there exists a close
connection between the quantum fluctuations and the averaged
magnetoconductance effect; let us now turn to the analysis of the
mean magnetoconductance effect. \par
\medskip
{\bf II-3\nobreak\ The Mean Magnetoconductance Effect} \par
\smallskip
Positive magnetoconductance at low temperature in
insulating GaAs:Si was reported long ago $ [3-13-22]. $ Amongst the
models which have been proposed, Spivak and Shklovskii $ [8-21] $
predict at the macroscopic limit, that:
$$ {\rm ln} ({\sigma (H\gg H_c) \over \sigma (0)}) \simeq  1 \eqno (11) $$
$ (H_c $ is given by $ {\pi c\hbar \over r^{3/2}_0\xi^{ 1/2}e}), $ which
compares very well with
numerical simulation $ [8]. $ \par
Zhao et al. $ [10] $ argue that simulations performed within
the same framework of FDP analysis but on larger samples, show no
saturation of the magnetoconductance in the limit of very large
quantum coherent sample (i.e. very low temperature). Moreover
they give a universal estimation:
$$ {\rm ln} ({\sigma (H) \over \sigma (0)})\simeq 0.1{r_0 \over L_H} {\rm \ \
\ where\ \ } \ L_H=({hc \over e}H)^{{-1 \over 2}} \eqno (12) $$
Their simulations corroborate the results obtained by Medina et
al. $ [11]. $ \par
As noted in the introduction, RMT predictions differ from the
FDP model because the positive magnetoconductance (in case of
negligible spin-orbit scattering) depends on $ {r_0 \over \xi} $ and not only
on $ r_0 $
(the predictions differ completely in case of strong spin-orbit
scattering). For instance at $ T=0 {\rm K} $ (to avoid the introduction of the
phase coherent hop and its magnetic field dependence):
$$ {\rm ln} ({g(H>H^\ast_ c) \over g(0)})\sim -{L \over \xi (H>H_c)}+{L \over
\xi (0)}={L \over 2\xi (0)}=-{1 \over 2} {\rm ln} (g(0)) \eqno (13) $$
if $ \xi (H>H^\ast_ c)= 2 \xi (0) $ $ [12] $ (quasi 1D RMT result; $ H^\ast_ c
$ is given by $ H^\ast_ c\xi^ 2\simeq  {hc \over e}). $
At finite temperature, the expression is less simple because $ r_0 $
depends on $ H $ via $ \xi (H) $ $ [12]. $ Nevertheless, {\sl the mean
magnetoconductance
is very sensitive to the mean conductance in zero field\/} in this RMT
approach. \par
Figure 12 shows the mean magnetoconductance effect between $ H=0 $
and $ H=2.5 $\nobreak\ Teslas in the strongly localized regime. The zero field
mean
conductance -\nobreak\ experimentally the smoothed conductance after numerical
averaging of the fluctuations in $ V_G\ - $ varies over 3 orders of
magnitude. Nevertheless $ < {\rm ln} ({g(H=2.5 {\rm T} ) \over g(0)})> $ is
roughly unchanged and
approximately equal to 1. The averaging over the {\sl whole \/}range $ V_G $
gives
$ {\rm ln} ({g(2.5 {\rm T} ) \over g(0)})\simeq 1. $ This is just the
prediction of NSS $ [8-21] $ (the
averaging needed for this prediction is obtained by smoothing in $ V_G $
which extends over several fluctuations). We note that it is also in
good accordance with the result of Zhao et al. $ [11] $ if we suppose that
$ r_0\simeq 160 {\rm nm,} $ a realistic value in our experiment, is roughly
insensitive to $ <g(0)>. $ However we are not able to test their analytical
universal result. In any case, the insensitivity of the mean
magnetoconductance to the mean conductance value stresses the fact that
FDP approaches are more adapted than RMT approaches in this regime. \par
As one approaches the Anderson transition, the mean
magnetoconductance tends smoothly to the weak antilocalization
contribution in the diffusive regime. Contrarily to the strongly
insulating regime where the mean magnetoconductance and the
conductance fluctuations are of the same order of magnitude, near
the transition the mean magnetoconductance becomes much larger
than the fluctuations. The analysis in the barely localized regime
in terms of changes of the localization length is restricted
because of the small range of conductance where this regime
occurs. Nevertheless the observed magnetoconductance is compatible
with a small increase of $ \xi $ (for instance $ \xi (2.5 {\rm T} )=1.3\xi (0)
$ for
$ T_0\simeq 2 {\rm K} ), $ as predicted by RMT approach $ [9]: $
$$ \xi =(\beta N+2-\beta )l \eqno (14) $$
where $ N $ is the number of transverse channels and $ \beta $ is one or two
respectively in absence or in presence of applied magnetic field.
In our experiment $ N $ is close to one, so that the crossover from
$ \beta =1 $
to $ \beta =2 $ does not imply a doubling of $ \xi , $ as predicted in the
macroscopic limit $ (N\mapsto \infty ). $ \par
\medskip
\vfill\eject
\noindent{\bf CONCLUSION} \par
\smallskip
The initial aim of this work was to gain more insight into
quantum interferences phenomena in a mesoscopic, disordered
insulator. We have studied a small wire where enormous
reproducible conductance fluctuations versus the Fermi energy of
electrons or versus applied magnetic field are observed at very
low temperature. \par
The fluctuation versus Fermi energy results from an
interplay between geometrical incoherent and quantum mechanically
coherent mesoscopic effects. The fluctuation versus magnetic
field,
on the contrary, is purely due to interference effects. \par
We are able to distinguish two insulating regimes. When
$ \xi $
is comparable to the distance between impurities, the observed
non-ergodicity and the analysis of the mean magnetoconductance
$ (< {\rm ln} ({g(H>1) \over g(0)} )>\simeq 1) $ indicates that interferences
between Time
Reversal conjugated loops are not essential to describe the
properties of the conductance distribution. On the other hand,
close to the Anderson transition, the localization radius includes
many impurity sites. Unfortunately, this critical regime is narrow
in our sample, so that a precise comparison with the predictions
of the RMT approach is not available, except the important fact
that the fluctuation is ergodic. We believe that, in the
experiment, we do not mistake this critical insulating regime for
the critical diffusive regime near the Anderson transition. In any
case, as far as a finite temperature experiment can determine the
critical transition point, the ergodicity holds beyond the
diffusive regime. \par
In the variable range hopping regime of our 1D sample deep
enough in the insulating regime, a theory where only one
elementary long hop dominates the resistance gives a good
quantitative prediction for the fluctuation versus energy. \par
Finally, the study of mesoscopic insulators with a larger
disorder, like amorphous alloys, will give us more insight into
the critical Anderson insulating phase. \par
\medskip
\noindent{\bf AKNOWLEDGEMENTS} \par
\smallskip
We would particularly like to thank J-L. Pichard and J-P
Bouchaud for their constant collaboration on this work. We are
also very grateful to B. Etienne for providing the GaAs:Si layer
and to R. Tourbot for his technical support. We also acknowledge
fruitful discussions with R. Jalabert, K. Slevin and B.I.
Shklovskii. \par
\vfill\eject
\centerline{{\bf REFERENCES}}
\noindent \vglue 2truecm \par
\smallskip
\noindent \item {$\lbrack$1$\rbrack$}S. Washburn in Quantum Coherence in
Mesoscopic Systems Kramer
ed. NATO ASI Serie 1991 Plenum Press, and ref. therein. \par
\smallskip
\noindent \item {$\lbrack$2$\rbrack$}A. Fowler, J Wainer and R. Webb in
Hopping Transport in Solids
Pollack and Shklovskii ed. Elsevier Science Publisher 1991. \par
\smallskip
\noindent \item {$\lbrack$3$\rbrack$}A. Orlov, A. Savchenko and A. Koslov,
Solid State Comm. {\bf 72},{\bf\ }743
(1989);
and E.I. Laiko, A.O. Orlov, A.K. Savchenko, E.A. Ilyichev and E.A.
Poltoratsky Zh Eksp Teor Fiz. {\bf 93}, 2204 (1987) (Sov. Phys. JETP
{\bf 66}, 1264 (1987)). \par
\smallskip
\noindent \item {$\lbrack$4$\rbrack$}N.F. Mott J. non-crystalline Solids {\bf
1},{\bf\ }1 (1968). \par
\smallskip
\noindent \item {$\lbrack$5$\rbrack$}P.A. Lee Phys. Rev. Lett. {\bf 53},{\bf\
}2042 (1984). \par
\smallskip
\noindent \item {$\lbrack$6$\rbrack$}M.E. Raikh and I.M. Ruzin, in \lq\lq
Mesoscopic Phenomena in
Solids\rq\rq , B.L. Altshuler, P.A. Lee and R.A. Webb editors, North
Holland 1991. \par
\smallskip
\noindent \item {$\lbrack$7$\rbrack$}V. Ambegaokar, B. Halperin and J.S.
Langer, Phys. Rev. {\bf B4},{\bf\ }2612
(1971). \par
\smallskip
\noindent \item {$\lbrack$8$\rbrack$}V.L. Nguen, B.Z. Spivak and B.I.
Shklovskii Piz'ma Zh Eksp Teor
Fiz. {\bf 43},{\bf\ }35 (1986) (JETP Lett. {\bf 43},{\bf\ }44 (1986)). \par
\smallskip
\noindent \item {$\lbrack$9$\rbrack$}J.L. Pichard in Quantum Coherence in
Mesoscopic Systems Kramer
ed. NATO ASI Serie Plenum Press 1991, and ref. therein. \par
\smallskip
\noindent \item {$\lbrack$10$\rbrack$}H.L. Zhao, B.Z. Spivak, M.P. Gelfand and
S. Feng, Phys. Rev.
{\bf B44, }10760 (1991). \par
\smallskip
\noindent \item {$\lbrack$11$\rbrack$}E. Medina and M. Kardar, Phys. Rev.
Lett.{\bf\ 66},{\bf\ }3187 (1991); and
E. Medina, M. Kardar, Y. Shapir and X.R. Wang, Phys. Rev. Lett.
{\bf 64},{\bf\ }1816 (1990). \par
\smallskip
\noindent \item {$\lbrack$12$\rbrack$}J.L. Pichard, M. Sanquer, K. Slevin and
P. Debray, Phys. Rev.
Lett. {\bf 65}, 1812 (1990). \par
\smallskip
\noindent \item {$\lbrack$13$\rbrack$}J.P. Bouchaud, J. Phys. $\Irm$ France
{\bf 1, }985 (1991). \par
\smallskip
\noindent \item {$\lbrack$14$\rbrack$}S. Feng and J.L. Pichard, Phys. Rev.
Lett. {\bf 67},{\bf\ }753 (1991). \par
\smallskip
\noindent \item {$\lbrack$15$\rbrack$}F. Ladieu and J.P. Bouchaud, Companion
Paper, submitted to
J. Physique 1. \par
\smallskip
\noindent \item {$\lbrack$16$\rbrack$}B.L. Altshuler and A.G. Aronov, JETP
Lett. {\bf 33}, 499 (1981). \par
\smallskip
\noindent \item {$\lbrack$17$\rbrack$}D. Mailly and M. Sanquer, J. Phys. 1
France {\bf 2},{\bf\ }357 (1992). \par
\smallskip
\noindent \item {$\lbrack$18$\rbrack$}J. Kurkijarvi, Phys. Rev. {\bf B8}, 922
(1973). \par
\smallskip
\noindent \item {$\lbrack$19$\rbrack$}A.D. Stone and P.A. Lee, Phys. Rev.
Lett. {\bf 54}, 1196 (1985). \par
\smallskip
\noindent \item {$\lbrack$20$\rbrack$}Y. Avishai and J.L. Pichard and K.A.
Muttalib, \lq\lq Quantum
Transmission in Disordered Insulators: Random Matrix Theory and
Transverse Localization\rq\rq , submitted to J. Physique 1. \par
\smallskip
\noindent \item {$\lbrack$21$\rbrack$}B.I. Shklovskii and A.L. Efros,
Electronic Properties of
Doped Semiconductors. Springer (1984). \par
\smallskip
\noindent \item {$\lbrack$22$\rbrack$}F. Tremblay, M. Pepper, D. Ritchie, D.C.
Peacock, J.E.F. Frost
and G.A.C. Jones, Phys. Rev. {\bf B39}, 8059 (1989). \par
\smallskip
\vfill\eject
\centerline{{\bf FIGURES CAPTIONS}}
\noindent \vglue 2truecm \par
\noindent \item {\underbar{Figure\nobreak\ 1:}}
The I-V Characteristic at $ T=91 {\rm mK} $ for a typical low conductance.
The solid line is a fit by an sh function (see the text). \par
\smallskip
\noindent \item {\underbar{Figure\nobreak\ 2:}}
Magnetoconductance (in units of $ {e^2 \over h}) $ at $ T=100 {\rm mK} $ for
large
positive $ V_G=+1.8V $ (diffusive regime). The solid line is the 1D
Weak Localization fit. The vertical bar is the UCF estimation. \par
\smallskip
\noindent \item {\underbar{Figure\nobreak\ 3:}}
The conductance (in quantum units) as function of $ V_G $ in the
diffusive regime. The vertical bar is the UCF estimation. \par
\smallskip
\noindent \item {\underbar{Figure\nobreak\ 4:}}
A: $ {\rm ln} (g) $ (in quantum units) versus $ T^{-{1 \over 2}} $ for various
$ V_G. $ The $ T_0 $
parameter values for the extremal curves are indicated. \par
\noindent \item {\nobreak\ \nobreak\ \nobreak\ \nobreak\ \nobreak\ \nobreak\
\nobreak\ \nobreak\ \nobreak\ }B: $ {\rm ln} (g) $ versus $ V_G $ at various
temperatures between $ T\simeq 1 {\rm K} $ and
$ T\simeq 70 {\rm mK} . $
The range of $ V_G $ corresponds to the curves at the bottom of
figure 4A. \par
\smallskip
\noindent \item {\underbar{Figure\nobreak\ 5:}}
A: Conductance in quantum units versus the gate voltage at $ T=4.2 {\rm K.} $
\par
\noindent \item {\nobreak\ \nobreak\ \nobreak\ \nobreak\ \nobreak\ \nobreak\
\nobreak\ \nobreak\ \nobreak\ }B: The same at $ T=100 {\rm mK.} $ \par
\smallskip
\noindent \item {\underbar{Figure\nobreak\ 6:}}
Figure 5B in a semi-logarithmic plot. The arrows indicates the
estimated Anderson transition and the barely and strongly
insulating regimes where the magnetic field dependence has been
precisely studied (see figure 8). \par
\smallskip
\noindent \item {\underbar{Figure\nobreak\ 7:}}
$ \delta {\rm ln} (R) $ versus $ < {\rm ln} \ R> $ in quantum units at $ T=100
{\rm mK.} $ Two experiments
are represented to improve the statistics. $ < {\rm ln\ } g> $ is obtained
after
smoothing of the experimental curves $ g(V_G). $ Dotted lines are the
prediction of reference $ [15]. $ Note nevertheless the tendency of
$\delta$lnR to saturate at $ \simeq 1 $ for high resistances. \par
\smallskip
\noindent \item {\underbar{Figure\nobreak\ 8:}}
Conductance as function of $ V_G $ and $ H $ in a 3D-Plot. The $ V_G $ ranges
are indicated on figure 6. \par
\noindent \item {\nobreak\ \nobreak\ \nobreak\ \nobreak\ \nobreak\ \nobreak\
\nobreak\ \nobreak\ \nobreak\ }A: Low Conductances in a linear scale \par
\noindent \item {\nobreak\ \nobreak\ \nobreak\ \nobreak\ \nobreak\ \nobreak\
\nobreak\ \nobreak\ \nobreak\ }B: Low Conductances in a logarithmic scale \par
\noindent \item {\nobreak\ \nobreak\ \nobreak\ \nobreak\ \nobreak\ \nobreak\
\nobreak\ \nobreak\ \nobreak\ }C: Moderate conductances in a linear scale \par
\noindent \item {\nobreak\ \nobreak\ \nobreak\ \nobreak\ \nobreak\ \nobreak\
\nobreak\ \nobreak\ \nobreak\ }D: Moderate Conductances in a logarithmic scale
\par
\smallskip
\noindent \item {\underbar{Figure\nobreak\ 9:}}
Contour plots of figures 8B and 8D. The magnetic field does not
decorrelate the conductance pattern versus gate voltage in the
strongly localized regime (9A). On the contrary the situation is
ergodic in the barely localized regime (9B). \par
\smallskip
\noindent \item {\underbar{Figure\nobreak\ 10:}}
The smoothed conductance for $ H=0 $ and $ H=2.5 $ Teslas in the strongly
localized regime. The observed mean magnetoconductance effect is
roughly insensitive to the conductance value (at $ H=0) $ and
fluctuates around:$ {\rm ln} ({g(H=2.5 {\rm T} ) \over g(H=0)}) \simeq  1, $
the mean
magnetoconductance value after averaging over the whole range of
$ V_G. $ \par
\smallskip
\bye
\listrefs
\draftend
\end